\documentclass[twocolumn,preprintnumbers,amsmath,amssymb,superscriptaddress,showpacs]{revtex4}

\usepackage{graphicx}

\usepackage[latin1]{inputenc}
\usepackage{upgreek}

\newcommand{\ie}{\textsl{i.e.}}

\begin{document}

\title{Disordered, strongly scattering porous materials as miniature multipass gas cells}

\author{Tomas Svensson}
\affiliation{Division of Atomic Physics, Department of Physics, Lund University, Box 118, 221 00 Lund, Sweden}
\author{Erik Adolfsson}
\affiliation{Ceramic Materials, SWEREA IVF, 431 83 M\"{o}lndal, Sweden}
\author{M\"arta Lewander}
\author{Can T. Xu}
\author{Sune Svanberg}
\affiliation{Division of Atomic Physics, Department of Physics, Lund University, Box 118, 221 00 Lund, Sweden}

\begin{abstract}

Spectroscopic gas sensing is both a commercial success and a rapidly advancing scientific field. Throughout the years, massive efforts have been directed towards improving detection limits by achieving long interaction pathlengths. Prominent examples include the use of conventional multipass gas cells
\cite{White1942_JOptSocAm,Herriott1964_ApplOpt,McManus1995_ApplOpt,Chernin1991_ApplOpt}, sophisticated high-finesse cavities
\cite{OKeefe1988_RevSciInstrum,Engeln1998_RevSciInstrum,Ye1998_JOSAB,Bernhardt2010_NatPhoton}, gas-filled holey fibers
\cite{Ritari2004_OptExpress,Ghosh2005_PhysRevLett,Hoo2010_IEEEPhotonicTechL}, integrating spheres \cite{Tranchart1996_ApplOpt,Hawe2007_MeasSciTechnol,Masiyano2010_ApplPhysB}, and diffusive reflectors \cite{Hodgkinson2010_ApplPhysB,Chen2010_ApplPhysB}. Despite this rich flora of approaches, there is a continuous struggle to reduce size, gas volume, cost and alignment complexity. Here, we show that extreme light scattering in porous materials can be used to realise miniature gas cells. Near-infrared transmission through a 7 mm zirconia (ZrO$_2$) sample with a 49\% porosity and subwavelength pore structure (on the order of 100 nm) gives rise to an effective gas interaction pathlength above 5 meters, an enhancement corresponding to 750 passes through a conventional multipass cell. This essentially different approach to pathlength enhancement opens a new route to compact, alignment-free and low-cost optical sensor systems.

\end{abstract}



\maketitle

\begin{figure}
  \includegraphics[]{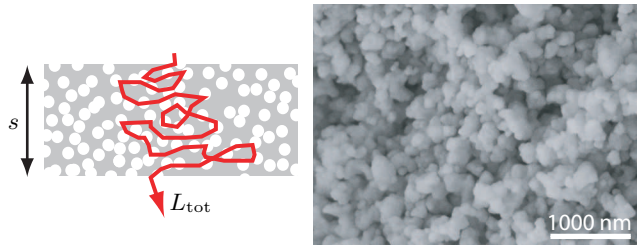}
  \caption{Illustration of porous media as gas cells (left), and a scanning electron microscope image of a utilized nanoporous zirconia with 115 nm pores (right). When light propagates through a strongly scattering and porous material of thickness $s$, the average pathlength of transmitted light ($L_\textrm{tot}$) will greatly exceed $s$. The pathlength through gas ($L$) is approximately proportional to $L_\textrm{tot}$ and the porosity $\phi$. The enhancement of the gas interaction pathlength may be defined as $N=L/s$. \label{fig:Principle}}
\end{figure}

Disordered, strongly scattering materials are continuously finding new applications in photonics. Besides constituting the physical system for fundamental investigation of Anderson localization of light \cite{Wiersma1997_Nature,Storzer2006_PhysRevLett} and random lasing \cite{Cao1999_PhysRevLett,Wiersma2008_Nature}, strong turbidity can also be used for, e.g., light trapping in solar cells \cite{Garnett2010_NanoLett} and focussing beyond the diffraction limit \cite{Vellekoop2010_NatPhoton}. The use of porous and strongly scattering media as efficient multipass gas cells, as described here, is an exciting addition to this diverse field. The main idea is to inject light into the porous material and collect diffusively transmitted light at some distance from the injection spot. Multiple scattering will force photon pathlengths to greatly exceed the physical source-detector separation (Figure \ref{fig:Principle}), and this random multipass effect will result in enhanced gas absorption imprints. As shown in previous work on spectroscopy of gases in porous media \cite{Sjoholm2001_OptLett}, absorption related to the solid is easily distinguished from gas phase absorption due to the great difference in spectral sharpness. Here, by utilizing disordered and highly porous materials based on low-absorption and high-refractive index ceramic materials such as alumina (Al$_2$O$_3$, refractive index $n=1.76$ at 760 nm), zirconia (ZrO$_2$, $n=2.14$) and titania (TiO$_2$, $n=2.49$), we report on remarkable pathlength enhancements. The strong scattering of nanoporous alumina and its possible application to spectroscopic gas sensing, was brought to attention in recent work on spectroscopy of confined gases and wall collision line broadening \cite{Svensson2010_ApplPhysLett}. Manufacturing and use of materials with much higher scattering efficiency, as presented here, represent dramatic improvements in terms of pathlength enhancement and gas cell potential.

We have manufactured and studied five different porous sintered ceramic materials (one titania, two alumina, and two zirconia materials). The pore structure has been investigated using mercury intrusion porosimetry \cite{Rouquerol1994_PureApplChem}, and all materials have narrow, well defined pore size distributions. The titania is nanoporous (sintered Kronos 1001) and has a 42\% porosity and pores in the 79$\pm$10 nm range (width here defined as the half width at half max, HWHM, of the pore size distribution). One alumina is nanoporous with 69$\pm$8 nm pores and a 35\% porosity (AKP30, Sumitomo, Japan), and the other is macroporous with 3.7$\pm$0.4 $\upmu$m pores and a 34\% porosity (made to 92.5\% from AA10, Sumitomo, the rest being a 40 nm silica powder used as a binder). The zirconia materials are nanoporous and both have a 49\% porosity, pores being 43$\pm$6 nm (TZ3YBE, Tosho, Japan) and 115$\pm$15 nm, respectively (TZ3YSBE, Tosho, Japan).

\begin{figure}
  \includegraphics[]{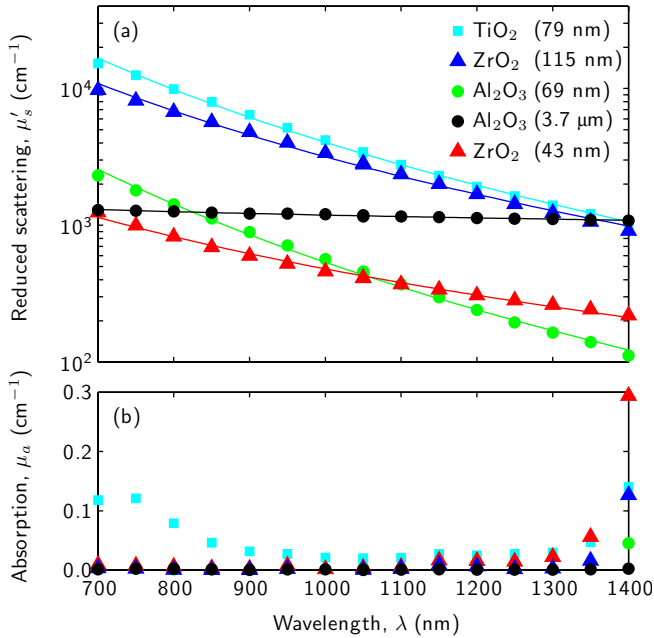}
  \caption{Optical properties of the ceramic materials (measured using PTOFS). Reduced scattering coefficients are shown in (a), and absorption coefficients in (b). Solid lines in (a) show fitted $a\times(\lambda/\mu m)^{-b}$ curves
  and give information on scattering decay, being $4045\times\lambda^{-4.0}$ for the titania, $537\times\lambda^{-4.4}$ and $1187\times\lambda^{-0.3}$ for the aluminas, and $3167\times\lambda^{-3.5}$ and $480\times\lambda^{-2.4}$ for the zirconias, respectively. \label{fig:OptProp}}
\end{figure}

Optical properties (scattering and absorption) of these materials have been measured in the 700-1400 nm range by employing photon time-of-flight spectroscopy (PTOFS) \cite{Svensson2009_RevSciInstrum}. The measured reduced scattering coefficients ($\mu_s'$) are presented in Figure \ref{fig:OptProp}a, revealing a strong dependence on refractive index and pore size. The scattering of the two strongest scatterers even approach that of the materials used in the quest for Anderson localization of light \cite{Wiersma1997_Nature,Storzer2006_PhysRevLett}. At 700 nm, the transport mean free path of photons ($l^*=1/\mu_s'$) are about 0.7 $\upmu$m and 1.0 $\upmu$m for the titania and 115 nm zirconia, respectively (the shortest photon mean free path observed, published in connection with work on Anderson localization, is about 0.2 $\upmu$m \cite{Wiersma1997_Nature}). It is also interesting to note that the materials with pore size smaller than the wavelength exhibit a $\lambda^{-4}$ Rayleigh-type scattering dependency. The 40 nm zirconia is partly an exception, indicating that scattering due to collective heterogeneity appears to have a strength comparable to that of individual pores. The macroporous alumina, on the other hand, shows that a heterogeneity larger than the wavelength can be used to maintain high scattering over a large spectral range.

The estimated absorption of the materials are shown in Figure \ref{fig:OptProp}b. As expected for these ceramic materials, the absorption is low in the near-infrared range. The titania, however, exhibit a fairly strong absorption in the short wavelength region, being a tail of strong absorption in the ultraviolet. It should also be noted that all materials, except the macroporous alumina, exhibit significant absorption at 1400 nm related to adsorbed water.

\begin{figure}
  \includegraphics[]{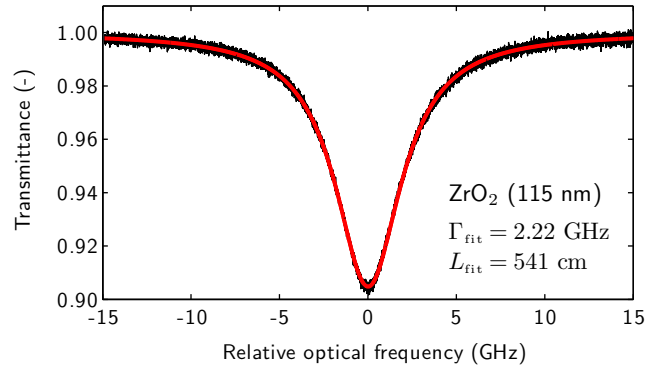}
  \caption{Oxygen (O$_2$) absorption for light transmitted through the zirconia material with 115 nm pores. The baseline-corrected experimental gas spectrum (black, noisy) is shown together with a fitted Lorentzian lineshape (red, smooth). Fitted pathlength through gas and line HWHM are stated. The acquisition time was 10 s, and the detected power was about 10 nW (0.003\% transmission). \label{fig:TDLAS_ZrO2_grov}}
\end{figure}

The potential of the materials as spectroscopic multipass gas cells are demonstrated in high-resolution near-infrared (760.654 nm, vacuum wavelength of transition) laser absorption spectroscopy of molecular oxygen. At atmospheric conditions, the line has a peak absorption coefficient of $2.7\times10^{-4}$ cm$^{-1}$ and a linewidth of 1.6 GHz HWHM.
\begin{table*}
  \centering
  \caption{Results from near-infrared (760.654 nm) spectroscopy of molecular oxygen in sintered ceramic materials. The table states material thickness ($s$), photon transport mean free path ($l^*$), porosity ($\phi$), pore size ($d$), detected transmission ($T$), spectroscopic linewidth ($\Gamma$), pathlength through gas ($L$), and pathlength enhancement ($N=L/s$). Sample diameters were in all cases 12-14 mm.}\label{tab:data}
  \setlength{\tabcolsep}{8pt}
  \begin{tabular}{lllllllll}
    \hline
    Material    & $s$ (mm)    & $l^*$ ($\upmu$m)  & $\phi$ (\%)      & $d$ (nm)       & $T$ (\%) & $\Gamma$ (GHz)             & $L$ (cm)           & $N=L/s$ (-)   \\
    \hline
    ZrO$_2$     & 7.2         & 1.3               & 48.8             & 115$\pm$15     & 0.003    & 2.218$\pm$0.009            & 541$\pm$4          & $\simeq$750   \\
    Al$_2$O$_3$ & 5.5         & 5.8               & 34.5             & 69$\pm$8       & 0.02     & 2.36$\pm$0.02              & 83.3$\pm$0.8       & $\simeq$150   \\
    TiO$_2$     & 1.4         & 0.8               & 42.4             & 79$\pm$10      & 0.001    & 2.17$\pm$0.09              & 19$\pm$2           & $\simeq$135   \\
    ZrO$_2$     & 7.0         & 10.4              & 48.6             & 43$\pm$6       & 0.05     & 2.75$\pm$0.06              & 86$\pm$4           & $\simeq$120   \\
    Al$_2$O$_3$ & 9.9         & 7.8               & 34.0             & 3700$\pm$400   & 0.02     & 1.616$\pm$0.003            & 59.5$\pm$0.2       & $\simeq$60    \\
    \hline
  \end{tabular}
\end{table*}
Spectroscopic results are presented in Table 1, and the experimental gas spectrum from the most potent material (the zirconia with 115 nm pores) is shown in Figure \ref{fig:TDLAS_ZrO2_grov}. The pathlength enhancement of this zirconia is impressive, and correspond to 750 passes through a conventional multipass cell. It is, however, important not to look only at the achieved pathlength enhancement. The average pathlength of light transmitted through a turbid material is proportional to the square of the thickness. This holds also for the pathlength through gas \cite{Somesfalean2002_ApplOpt}, \ie\ $L\propto s^2$, and the pathlength enhancement is therefore proportional to the material thickness, $N\propto s$. Instead, the suitability of a porous material as a random multipass gas cell is jointly determined by the scattering efficiency, the absorption of the solid, and the material porosity. Strong scattering and high porosity being ideal properties for this purpose, while absorption is detrimental both in terms of pathlength through gas and total transmission. Although the titania material exhibits the strongest scattering, the rather strong absorption at 760 nm makes it a less successful gas cell candidate  than the zirconia (at this wavelength). It is also important to realise that losses due to scattering and absorption sets a limit to the maximum possible thickness. Diffuse light propagation cause transmitted intensity to decrease with $1/s$, even if the absorption would be zero. For example, only 0.003\% of the light incident on the 7.2 mm thick zirconia reached the detector. For comparison, the strong scattering combined with absorption of the titania (at 760 nm), makes it difficult to boost the pathlength through gas by increasing sample thickness (only 0.001\% reaches through the 1.4 mm sample).

Another interesting aspect of the results in Table 1, not yet mentioned, is the observed linewidths. The linewidth of oxygen within the macroporous alumina agrees well with the width expected for oxygen at atmospheric pressure, \ie\ 1.6 GHz. In great contrast, oxygen confined in the nanoporous materials exhibits significantly broadened absorption lines. This effect was only recently described \cite{Svensson2010_ApplPhysLett,Svensson2010_OptExpress}, and is related to the fact that wall collisions no longer are negligible. For oxygen at ambient conditions, the mean free path between intermolecular collisions is about 60 nm. Since these collisions are the dominating source of line broadening at atmospheric conditions, it is not surprising that confinement in nano-cavities gives rise to additional collisional broadening. Furthermore, the clear negative correlation between pore size and line broadening clearly shows the potential of wall collision broadening as means to measure pore size, as suggested in Ref. \cite{Svensson2010_ApplPhysLett}.

While the above presented data are obtained for spectroscopy at 760 nm, similar results can be reached for a wide spectral range. The absorption of the ceramic materials is fairly low up to 4 $\upmu$m, and by accompanying a longer wavelength with a scaling of pore size and sample thickness, the scattering efficiency and pathlength enhancement can be maintained. This means that porous ceramics may be used as spectroscopic gas cells for numerous important gases, including methane (CH$_4$), carbon dioxide (CO$_2$), carbon monoxide (CO), nitrous oxide (NO), nitrogen dioxide (NO$_2$), ammonia (NH$_3$) and water (H$_2$O). An issue that requires further investigation is, however, how the response time of porous gas cells varies between gases, as well with pore size and cell geometry. While relatively inert gases, such as oxygen, have been found to move rapidly in and out of nanoporous alumina \cite{Svensson2010_ApplPhysLett,Svensson2010_OptExpress}, the exchange of water vapor (known to be a "sticky" gas) is fairly slow \cite{Svensson2010_OptExpress}.

To summarize, we have shown that strongly scattering porous materials can function as miniature gas cells. Important advantages of such gas cells are their small size, the small gas volume needed (despite strong pathlength enhancement), and their alignment-free nature. The most obvious disadvantage is that strong scattering results in only a small fraction of incident light reaching the detector. In the case of laser spectroscopy, another problem is speckle interference noise \cite{Svensson2008_OptLett,Svensson2010_OptExpress}. However, the topic is new and significant improvements regarding detection schemes and gas cell design can be anticipated. For example, other configurations than the pure transmission geometry (as adopted here) can be explored. This includes designs where lost light is reinjected into the material, where light is launched inside the porous gas cell, or where a reflection-type geometry is employed. Porous gas cells can also be manufactured from other crystal forms of the ceramics or other low-absorption materials. Finally, the pore structure (porosity and pore size) can be optimized to yield stronger scattering \cite{Rivas2002_ApplPhysLett}, allowing smaller gas cells while maintaining or increasing $L$. In particular, materials with pore sizes in the 100-3000 nm range need to be explored.


\section*{METHODS}

\textbf{Sintering} Titania and zirconia materials were sintered for 2 hours at 900$^\circ$C. The nanoporous alumina was sintered for 10 minutes at 1000$^\circ$C, and the macroporous alumina for 45 minutes at 1400$^\circ$C.

\textbf{High-resolution spectroscopy} is based on tunable diode laser absorption spectroscopy (TDLAS), and the utilized instrument is described in detail in Ref. \cite{Svensson2008_APB}. Optical interference noise, an important limitation in laser spectroscopy of gases in turbid porous media, is suppressed by means of laser beam dithering \cite{Svensson2008_OptLett}. Briefly, light from a 0.3 mW VCSEL diode laser is injected into the pillbox-shaped ceramic samples, and diffuse transmission is detected by a $5.6\times 5.6$ mm$^2$ large-area photodiode (no alignment needed). The laser is tuned over the R9Q10 absorption line at 760.654 nm. This transition has a line strength of about $2.6\times10^{-13}$ cm$^2$Hz, which at atmospheric conditions (1 atm, 293 K, 20.9\% O$_2$) gives rise to a peak absorption coefficient of about $2.7\times10^{-4}$ cm$^{-1}$. The corresponding linewidth is about 1.6 GHz HWHM, originating from a 1.4 GHz pressure broadening combined with a 0.4 GHz Doppler broadening.

\textbf{Photon time-of-flight spectroscopy} is conducted using a tunable system based on a fiber laser combined with super-continuum generation, acousto-optical tunable filters, and time-correlated single photon counting
\cite{Svensson2009_RevSciInstrum}. Picosecond light pulses are injected into thin, polished-down versions of the ceramic materials, and the temporal shape of the transmitted pulses allows, via diffusion theory \cite{Contini1997_ApplOpt}, assessment of absorption and reduced scattering coefficients. Since effective medium theories for strongly scattering materials are not fully developed \cite{Schuurmans1999_Science}, the evaluation is based on a volume-averaged refractive index of the porous materials (using appropriate Sellmeier equations for the solids). The volume-averaging approach limits accuracy, since light in porous media tends to interact predominantly with the solid \cite{Svensson2010_OptLett}.

\section*{ACKNOWLEDGEMENTS}

This work was funded by the Swedish Research Council. T.S. gratefully acknowledge MSc. Erik Alerstam for important assistance in PTOFS experiments, former colleagues Dr. Mats Andersson and Dr. Lars Rippe for fruitful cooperation in development of the TDLAS instrumentation, as well as Dr. Dmitry Khoptyar and Prof. Stefan Andersson-Engels for general collaboration on PTOFS. Karin Lindqvist at SWEREA IVF is acknowledged for manufacturing the alumina materials.

\section*{AUTHOR CONTRIBUTIONS}

T.S. initiated the project, was the principal investigator, and drafted the manuscript. E.A. manufactured zirconia and titania materials. M.L. and C.X. assisted in TDLAS measurements. S.S. was instrumental in early concept developments. All authors approved the final manuscript version. Correspondence should be addressed to T.S.~(email: tomas.svensson@fysik.lth.se).


\begin{thebibliography}{10}
\expandafter\ifx\csname url\endcsname\relax
  \def\url#1{\texttt{#1}}\fi
\expandafter\ifx\csname urlprefix\endcsname\relax\def\urlprefix{URL }\fi
\providecommand{\bibinfo}[2]{#2}
\providecommand{\eprint}[2][]{\url{#2}}

\bibitem{White1942_JOptSocAm}
\bibinfo{author}{White, J.}
\newblock \bibinfo{title}{Long optical paths of large aperture}.
\newblock \emph{\bibinfo{journal}{J. Opt. Soc. Am.}}
  \textbf{\bibinfo{volume}{32}}, \bibinfo{pages}{285--288}
  (\bibinfo{year}{1942}).

\bibitem{Herriott1964_ApplOpt}
\bibinfo{author}{Herriott, D.}, \bibinfo{author}{Kompfner, R.} \&
  \bibinfo{author}{Kogelnik, H.}
\newblock \bibinfo{title}{Off-axis paths in spherical mirror interferometers}.
\newblock \emph{\bibinfo{journal}{Appl. Opt.}} \textbf{\bibinfo{volume}{3}},
  \bibinfo{pages}{523--526} (\bibinfo{year}{1964}).

\bibitem{McManus1995_ApplOpt}
\bibinfo{author}{McManus, J.~B.}, \bibinfo{author}{Kebabian, P.~L.} \&
  \bibinfo{author}{Zahniser, W.~S.}
\newblock \bibinfo{title}{Astigmatic mirror multipass absorption cells for
  long-path-length spectroscopy}.
\newblock \emph{\bibinfo{journal}{Appl. Opt.}} \textbf{\bibinfo{volume}{34}},
  \bibinfo{pages}{3336--3348} (\bibinfo{year}{1995}).

\bibitem{Chernin1991_ApplOpt}
\bibinfo{author}{Chernin, S.~M.} \& \bibinfo{author}{Barskaya, E.~G.}
\newblock \bibinfo{title}{Optical multipass matrix systems}.
\newblock \emph{\bibinfo{journal}{Appl. Opt.}} \textbf{\bibinfo{volume}{30}},
  \bibinfo{pages}{51--58} (\bibinfo{year}{1991}).

\bibitem{OKeefe1988_RevSciInstrum}
\bibinfo{author}{O'Keefe, A.} \& \bibinfo{author}{Deacon, D. A.~G.}
\newblock \bibinfo{title}{Cavity ring-down optical spectrometer for
  absorption-measurements using pulsed laser sources}.
\newblock \emph{\bibinfo{journal}{Rev. Sci. Instrum.}}
  \textbf{\bibinfo{volume}{59}}, \bibinfo{pages}{2544--2551}
  (\bibinfo{year}{1988}).

\bibitem{Engeln1998_RevSciInstrum}
\bibinfo{author}{Engeln, R.}, \bibinfo{author}{Berden, G.},
  \bibinfo{author}{Peeters, R.} \& \bibinfo{author}{Meijer, G.}
\newblock \bibinfo{title}{Cavity enhanced absorption and cavity enhanced
  magnetic rotation spectroscopy}.
\newblock \emph{\bibinfo{journal}{Rev. Sci. Instrum.}}
  \textbf{\bibinfo{volume}{69}}, \bibinfo{pages}{3763--3769}
  (\bibinfo{year}{1998}).

\bibitem{Ye1998_JOSAB}
\bibinfo{author}{Ye, J.}, \bibinfo{author}{Ma, L.~S.} \& \bibinfo{author}{Hall,
  J.~L.}
\newblock \bibinfo{title}{Ultrasensitive detections in atomic and molecular
  physics: demonstration in molecular overtone spectroscopy}.
\newblock \emph{\bibinfo{journal}{J. Opt. Soc. Am. B}}
  \textbf{\bibinfo{volume}{15}}, \bibinfo{pages}{6--15} (\bibinfo{year}{1998}).

\bibitem{Bernhardt2010_NatPhoton}
\bibinfo{author}{Bernhardt, B.}, \bibinfo{author}{Ozawa, A.},
  \bibinfo{author}{Jacquet, P.}, \bibinfo{author}{Jacquey, M.},
  \bibinfo{author}{Kobayashi, Y.}, \bibinfo{author}{Udem, T.},
  \bibinfo{author}{Holzwarth, R.}, \bibinfo{author}{Guelachvili, G.},
  \bibinfo{author}{Hansch, T.~W.} \& \bibinfo{author}{Picque, N.}
\newblock \bibinfo{title}{Cavity-enhanced dual-comb spectroscopy}.
\newblock \emph{\bibinfo{journal}{Nat. Photon.}}
  \textbf{\bibinfo{volume}{4}}, \bibinfo{pages}{55--57} (\bibinfo{year}{2010}).

\bibitem{Ritari2004_OptExpress}
\bibinfo{author}{Ritari, T.}, \bibinfo{author}{Tuominen, J.},
  \bibinfo{author}{Ludvigsen, H.}, \bibinfo{author}{Petersen, J.~C.},
  \bibinfo{author}{Sorensen, T.}, \bibinfo{author}{Hansen, T.~P.} \&
  \bibinfo{author}{Simonsen, H.~R.}
\newblock \bibinfo{title}{Gas sensing using air-guiding photonic bandgap
  fibers}.
\newblock \emph{\bibinfo{journal}{Opt. Express}} \textbf{\bibinfo{volume}{12}},
  \bibinfo{pages}{4080--4087} (\bibinfo{year}{2004}).

\bibitem{Ghosh2005_PhysRevLett}
\bibinfo{author}{Ghosh, S.}, \bibinfo{author}{Sharping, J.~E.},
  \bibinfo{author}{Ouzounov, D.~G.} \& \bibinfo{author}{Gaeta, A.~L.}
\newblock \bibinfo{title}{Resonant optical interactions with molecules confined
  in photonic band-gap fibers}.
\newblock \emph{\bibinfo{journal}{Phys. Rev. Lett.}}
  \textbf{\bibinfo{volume}{94}}, \bibinfo{pages}{093902}
  (\bibinfo{year}{2005}).

\bibitem{Hoo2010_IEEEPhotonicTechL}
\bibinfo{author}{Hoo, Y.~L.}, \bibinfo{author}{Liu, S.~J.},
  \bibinfo{author}{Ho, H.~L.} \& \bibinfo{author}{Jin, W.}
\newblock \bibinfo{title}{Fast response microstructured optical fiber methane
  sensor with multiple side-openings}.
\newblock \emph{\bibinfo{journal}{IEEE Photonic. Tech. L.}}
  \textbf{\bibinfo{volume}{22}}, \bibinfo{pages}{296--298}
  (\bibinfo{year}{2010}).

\bibitem{Tranchart1996_ApplOpt}
\bibinfo{author}{Tranchart, S.}, \bibinfo{author}{Bachir, I.} \&
  \bibinfo{author}{Destombes, J.}
\newblock \bibinfo{title}{Sensitive trace gas detection with near-infrared
  laser diodes and an integrating sphere}.
\newblock \emph{\bibinfo{journal}{Appl. Opt.}} \textbf{\bibinfo{volume}{35}},
  \bibinfo{pages}{7070--7074} (\bibinfo{year}{1996}).

\bibitem{Hawe2007_MeasSciTechnol}
\bibinfo{author}{Hawe, E.}, \bibinfo{author}{Chambers, P.},
  \bibinfo{author}{Fitzpatrick, C.} \& \bibinfo{author}{Lewis, E.}
\newblock \bibinfo{title}{{CO}$_2$ monitoring and detection using an
  integrating sphere as a multipass absorption cell}.
\newblock \emph{\bibinfo{journal}{Meas. Sci. Technol.}}
  \textbf{\bibinfo{volume}{18}}, \bibinfo{pages}{3187--3194}
  (\bibinfo{year}{2007}).

\bibitem{Masiyano2010_ApplPhysB}
\bibinfo{author}{Masiyano, D.}, \bibinfo{author}{Hodgkinson, J.} \&
  \bibinfo{author}{Tatam, R.}
\newblock \bibinfo{title}{Gas cells for tunable diode laser absorption
  spectroscopy employing optical diffusers. Part 2: Integrating spheres}.
\newblock \emph{\bibinfo{journal}{Appl. Phys. B}} \textbf{\bibinfo{volume}{100}},
  \bibinfo{pages}{303--312} (\bibinfo{year}{2010}).


\bibitem{Hodgkinson2010_ApplPhysB}
\bibinfo{author}{Hodgkinson, J.}, \bibinfo{author}{Masiyano, D.} \&
  \bibinfo{author}{Tatam, R.}
\newblock \bibinfo{title}{Gas cells for tunable diode laser absorption
  spectroscopy employing optical diffusers. Part 1: Single and dual pass
  cells}.
\newblock \emph{\bibinfo{journal}{Appl. Phys. B}} \textbf{\bibinfo{volume}{100}},
  \bibinfo{pages}{291--302} (\bibinfo{year}{2010}).

\bibitem{Chen2010_ApplPhysB}
\bibinfo{author}{Chen, J.}, \bibinfo{author}{Hangauer, A.},
  \bibinfo{author}{Strzoda, R.} \& \bibinfo{author}{Amann, M.-C.}
\newblock \bibinfo{title}{Laser spectroscopic oxygen sensor using diffuse
  reflector based optical cell and advanced signal processing}.
\newblock \emph{\bibinfo{journal}{Appl. Phys. B}} \textbf{\bibinfo{volume}{100}},
  \bibinfo{pages}{417--425} (\bibinfo{year}{2010}).


\bibitem{Wiersma1997_Nature}
\bibinfo{author}{Wiersma, D.~S.}, \bibinfo{author}{Bartolini, P.},
  \bibinfo{author}{Lagendijk, A.} \& \bibinfo{author}{Righini, R.}
\newblock \bibinfo{title}{Localization of light in a disordered medium}.
\newblock \emph{\bibinfo{journal}{Nature}} \textbf{\bibinfo{volume}{390}},
  \bibinfo{pages}{671--673} (\bibinfo{year}{1997}).

\bibitem{Storzer2006_PhysRevLett}
\bibinfo{author}{Störzer, M.}, \bibinfo{author}{Gross, P.},
  \bibinfo{author}{Aegerter, C.~M.} \& \bibinfo{author}{Maret, G.}
\newblock \bibinfo{title}{Observation of the critical regime near {A}nderson
  localization of light}.
\newblock \emph{\bibinfo{journal}{Phys. Rev. Lett.}}
  \textbf{\bibinfo{volume}{96}}, \bibinfo{pages}{063904}
  (\bibinfo{year}{2006}).

\bibitem{Cao1999_PhysRevLett}
\bibinfo{author}{Cao, H.}, \bibinfo{author}{Zhao, Y.~G.}, \bibinfo{author}{Ho,
  S.~T.}, \bibinfo{author}{Seelig, E.~W.}, \bibinfo{author}{Wang, Q.~H.} \&
  \bibinfo{author}{Chang, R. P.~H.}
\newblock \bibinfo{title}{Random laser action in semiconductor powder}.
\newblock \emph{\bibinfo{journal}{Phys. Rev. Lett.}}
  \textbf{\bibinfo{volume}{82}}, \bibinfo{pages}{2278--2281}
  (\bibinfo{year}{1999}).

\bibitem{Wiersma2008_Nature}
\bibinfo{author}{Wiersma, D.~S.}
\newblock \bibinfo{title}{The physics and applications of random lasers}.
\newblock \emph{\bibinfo{journal}{Nature Physics}}
  \textbf{\bibinfo{volume}{4}}, \bibinfo{pages}{359--367}
  (\bibinfo{year}{2008}).

\bibitem{Garnett2010_NanoLett}
\bibinfo{author}{Garnett, E.} \& \bibinfo{author}{Yang, P.~D.}
\newblock \bibinfo{title}{Light trapping in silicon nanowire solar cells}.
\newblock \emph{\bibinfo{journal}{Nano Lett.}} \textbf{\bibinfo{volume}{10}},
  \bibinfo{pages}{1082--1087} (\bibinfo{year}{2010}).

\bibitem{Vellekoop2010_NatPhoton}
\bibinfo{author}{Vellekoop, I.}, \bibinfo{author}{Lagendijk, A.} \&
  \bibinfo{author}{Mosk, A.~P.}
\newblock \bibinfo{title}{Exploiting disorder for perfect focusing}.
\newblock \emph{\bibinfo{journal}{Nat. Photon.}} \textbf{\bibinfo{volume}{4}},
  \bibinfo{pages}{320--322} (\bibinfo{year}{2010}).

\bibitem{Sjoholm2001_OptLett}
\bibinfo{author}{Sjöholm, M.}, \bibinfo{author}{Somesfalean, G.},
  \bibinfo{author}{Alnis, J.}, \bibinfo{author}{Andersson-Engels, S.} \&
  \bibinfo{author}{Svanberg, S.}
\newblock \bibinfo{title}{Analysis of gas dispersed in scattering media}.
\newblock \emph{\bibinfo{journal}{Opt. Lett.}} \textbf{\bibinfo{volume}{26}},
  \bibinfo{pages}{16--18} (\bibinfo{year}{2001}).

\bibitem{Svensson2010_ApplPhysLett}
\bibinfo{author}{Svensson, T.} \& \bibinfo{author}{Shen, Z.}
\newblock \bibinfo{title}{Laser spectroscopy of gas confined in nanoporous
  materials}.
\newblock \emph{\bibinfo{journal}{Appl. Phys. Lett.}}
  \textbf{\bibinfo{volume}{96}}, \bibinfo{pages}{021107}
  (\bibinfo{year}{2010}).

\bibitem{Rouquerol1994_PureApplChem}
\bibinfo{author}{Rouquerol, J.}, \bibinfo{author}{Avnir, D.},
  \bibinfo{author}{Fairbridge, C.~W.}, \bibinfo{author}{Everett, D.~H.},
  \bibinfo{author}{Haynes, J.~H.}, \bibinfo{author}{Pernicone, N.},
  \bibinfo{author}{Ramsay, J. D.~F.}, \bibinfo{author}{Sing, K. S.~W.} \&
  \bibinfo{author}{Unger, K.~K.}
\newblock \bibinfo{title}{Recommendations for the characterization of porous
  solids}.
\newblock \emph{\bibinfo{journal}{Pure Appl. Chem.}}
  \textbf{\bibinfo{volume}{66}}, \bibinfo{pages}{1739--1758}
  (\bibinfo{year}{1994}).

\bibitem{Svensson2009_RevSciInstrum}
\bibinfo{author}{Svensson, T.}, \bibinfo{author}{Alerstam, E.},
  \bibinfo{author}{Khoptyar, D.}, \bibinfo{author}{Johansson, J.},
  \bibinfo{author}{Folestad, S.} \& \bibinfo{author}{Andersson-Engels, S.}
\newblock \bibinfo{title}{Near infrared photon time-of-flight spectroscopy of
  turbid materials up to 1400 nm}.
\newblock \emph{\bibinfo{journal}{Rev. Sci. Instrum.}}
  \textbf{\bibinfo{volume}{80}}, \bibinfo{pages}{063105}
  (\bibinfo{year}{2009}).

\bibitem{Somesfalean2002_ApplOpt}
\bibinfo{author}{Somesfalean, G.}, \bibinfo{author}{Sjöholm, M.},
  \bibinfo{author}{Alnis, J.}, \bibinfo{author}{af~Klinteberg, C.},
  \bibinfo{author}{Andersson-Engels, S.} \& \bibinfo{author}{Svanberg, S.}
\newblock \bibinfo{title}{Concentration measurement of gas embedded in
  scattering media by employing absorption and time-resolved laser
  spectroscopy}.
\newblock \emph{\bibinfo{journal}{Appl. Opt.}} \textbf{\bibinfo{volume}{41}},
  \bibinfo{pages}{3538--3544} (\bibinfo{year}{2002}).

\bibitem{Svensson2010_OptExpress}
\bibinfo{author}{Svensson, T.}, \bibinfo{author}{Lewander, M.} \&
  \bibinfo{author}{Svanberg, S.}
\newblock \bibinfo{title}{Laser spectroscopy of water vapor confined in
  nanoporous alumina: wall collision line broadening and gas diffusion
  dynamics}.
\newblock \emph{\bibinfo{journal}{Opt. Express}}  (\bibinfo{year}{2010}).

\bibitem{Svensson2008_OptLett}
\bibinfo{author}{Svensson, T.}, \bibinfo{author}{Andersson, M.},
  \bibinfo{author}{Rippe, L.}, \bibinfo{author}{Johansson, J.},
  \bibinfo{author}{Folestad, S.} \& \bibinfo{author}{Andersson-Engels, S.}
\newblock \bibinfo{title}{High sensitivity gas spectroscopy of porous, highly
  scattering solids}.
\newblock \emph{\bibinfo{journal}{Opt. Lett.}} \textbf{\bibinfo{volume}{33}},
  \bibinfo{pages}{80--82} (\bibinfo{year}{2008}).

\bibitem{Rivas2002_ApplPhysLett}
\bibinfo{author}{Rivas, J.~G.}, \bibinfo{author}{Lagendijk, A}, \bibinfo{author}{Tjerkstra, R.~W.}, \bibinfo{author}{Vanmaekelberg, D}  \& \bibinfo{author}{Kelly, J.~J.}
\newblock \bibinfo{title}{Tunable photonic strength in porous GaP}.
\newblock \emph{\bibinfo{journal}{Appl. Phys. Lett.}}
  \textbf{\bibinfo{volume}{80}}, \bibinfo{pages}{4498--4500}
  (\bibinfo{year}{2002}).


\bibitem{Svensson2008_APB}
\bibinfo{author}{Svensson, T.}, \bibinfo{author}{Andersson, M.},
  \bibinfo{author}{Rippe, L.}, \bibinfo{author}{Svanberg, S.},
  \bibinfo{author}{Andersson-Engels, S.}, \bibinfo{author}{Johansson, J.} \&
  \bibinfo{author}{Folestad, S.}
\newblock \bibinfo{title}{{VCSEL}-based oxygen spectroscopy for structural
  analysis of pharmaceutical solids}.
\newblock \emph{\bibinfo{journal}{Appl. Phys. B}}
  \textbf{\bibinfo{volume}{90}}, \bibinfo{pages}{345--354}
  (\bibinfo{year}{2008}).

\bibitem{Contini1997_ApplOpt}
\bibinfo{author}{Contini, D.}, \bibinfo{author}{Martelli, F.} \&
  \bibinfo{author}{Zaccanti, G.}
\newblock \bibinfo{title}{Photon migration through a turbid slab described by a
  model based on diffusion approximation: I. Theory}.
\newblock \emph{\bibinfo{journal}{Appl. Opt.}} \textbf{\bibinfo{volume}{36}},
  \bibinfo{pages}{4587--4599} (\bibinfo{year}{1997}).

\bibitem{Schuurmans1999_Science}
\bibinfo{author}{Schuurmans, F. J.~P.}, \bibinfo{author}{Vanmaekelbergh, D.},
  \bibinfo{author}{van~de Lagemaat, J.} \& \bibinfo{author}{Lagendijk, A.}
\newblock \bibinfo{title}{Strongly photonic macroporous gallium phosphide
  networks}.
\newblock \emph{\bibinfo{journal}{Science}} \textbf{\bibinfo{volume}{284}},
  \bibinfo{pages}{141--143} (\bibinfo{year}{1999}).

\bibitem{Svensson2010_OptLett}
\bibinfo{author}{Svensson, T.}, \bibinfo{author}{Alerstam, E.},
  \bibinfo{author}{Johansson, J.} \& \bibinfo{author}{Andersson-Engels, S.}
\newblock \bibinfo{title}{Optical porosimetry and investigations of the
  porosity experienced by light interacting with porous media}.
\newblock \emph{\bibinfo{journal}{Opt. Lett.}} \textbf{\bibinfo{volume}{35}},
  \bibinfo{pages}{1740--1742} (\bibinfo{year}{2010}).

\end{thebibliography}
\end{document}